\documentstyle[11pt]{article}

\newcommand{\acknowledgements}{\ \\ \centerline{\bf Acknowledgements}\\}
\newcommand{\pacs}[1]{\ \\ \noindent PACS{ #1 } }

\begin{document}

\title{
Possibilities and Limitations of
Gaussian Closure Approximation\\
for Phase Ordering Dynamics.
}

\author{Chuck Yeung$^{a,c}$, Y. Oono$^{b}$ and A. Shinozaki$^{b,c}$
\\
        $^{a}$
        Department of Physics, University of Toronto,\\
        Toronto, Ontario
        M5S-1A7 CANADA\\
        $^{b}$
        Department of Physics, Materials Research Laboratory,\\
        and
        Beckman Institute,\\
        University of Illinois at Urbana-Champaign,\\
        Urbana, IL 61801 USA\\
        $^{c}$Department of Physics and Astronomy, University of
        Pittsburgh,\\
        Pittsburgh, PA 15260 USA
        }

\maketitle

\begin{abstract}

The nonlinear equations describing phase ordering dynamics can be
closed by assuming the existence of an underlying Gaussian stochastic
field which is nonlinearly related to the observable order parameter
field.   We discuss the relation between different implementations of
the Gaussian assumption and consider the limitations of this assumption
for phase ordering dynamics.  The fact that the different approaches
gives different results is a sign of the breakdown of the Gaussian
assumption.  We concentrate on the non-conserved order parameter case
but also touch on the conserved order parameter case.  We demonstrate
that the Gaussian assumption is fundamentally flawed in the latter case.

\end{abstract}


cond-mat/9305033

\newpage

\section{Introduction}

After a system is quenched from the disordered to the ordered phase,
domains of the ordered phases form and grow.  At late stages,
it is empirically known that phase ordering process obeys dynamical
scaling, {\em i.e.}, the spatial distribution of domains can be
described by a single time-dependent length, $L(t)$.  On this
length-scale, the phase ordering process depends only on a few general
features of the dynamics.  Due to the inherently nonlinear nature of
the dynamics, understanding the phase ordering process remains a
challenge \cite{GUNTON,FURUKAWA,BINDER}.  Analytic progress has been
confined to the case of ${\cal O}(n)$ component order-parameter in the
limit of large $n$ \cite{CONIGLIO,BRAYN,BRAYH}.  For $n \leq d$, where
$d$ is the spatial dimension, topological defects become important and
progress has been limited to dimensional analysis of the defect motion
\cite{FURUKAWA,ALLEN,KAWASAKI82} and methods by which the nonlinear
equations describing the dynamics are ``closed'' (closure
approximations)
\cite{LANGER73,LANGER75,KAWASAKI78,OHTA82,OONO88a,MAZENKO}.  For the
scalar order parameter case the simplest closure approximation is to
assume that the order parameter field $\phi({\bf r}, t)$ is a Gaussian
stochastic field \cite{LANGER73}.  This method fails, since, on the
scale of the characteristic domain size $L(t)$, $\phi({\bf r},t)$ is
effectively discontinuous.  Instead progress is achieved by assuming
that there exists an underlying Gaussian field which is nonlinearly
related to $\psi({\bf r}, t)$ \cite{OHTA82,OONO88a,MAZENKO}.

In this paper, we will explore
the reliability of closure approximations based on the assumption of an
underlying Gaussian stochastic field.  We first discuss phase ordering
dynamics without conservation of order parameter.  In particular we
will discuss the limitations of this assumption and the relation
between different implementations of the Gaussian closure.  One method
of introducing the underlying field is based on the dynamics of random
interfaces.  We will denote this scheme `I'.  It was first proposed by
Ohta, Jasnow and Kawasaki (OJK) \cite{OHTA82} for systems with
non-conserved order parameter (NCOP).  This method was later extended
by Ohta and Nozaki \cite{OHTAandNOZAKI} to systems with conserved order
parameter (COP).

A second manner of introducing the underlying field is by writing the
order parameter as $\psi({\bf r}, t) = f( u({\bf r}, t) )$, where $f$
is the interface profile.  We will call this method scheme `B' since it
relies on the full bulk dynamics.  Oono and Puri (OP) \cite{OONO88a}
introduced this scheme to remove inconsistencies in the original OJK
argument.  There are two manners in which the bulk closure scheme has
been applied. The first (scheme B$u$) obtains a closed equation for the
correlation function of $u({\bf r},t)$.  Oono and Puri demonstrated
that this provides reasonable results for the NCOP case
\cite{OONO88a}.  The second (scheme B$\psi$) constructs a closed
equation for the correlation function of $\psi({\bf r},t)$ using the
assumed Gaussian nature of the underlying field $u({\bf r},t)$.  Oono
applied the B$\psi$ closure to the COP case \cite{YOunpub} but found
that it did not yield the experimentally observed scaling, $L(t) \sim
t^{1/3}$.  Mazenko has applied the B$\psi$ closure to both the NCOP
\cite{MAZENKO} and COP cases \cite{MAZENKOCOP1,MAZENKOCOP2}.  For the
NCOP case, Mazenko found the B$\psi$ closure gives different results
from Ohta-Jasnow-Kawasaki and Oono-Puri but is also in reasonable
agreement with experiment\cite{MAZENKO}.  For the COP case, Mazenko
found that $L(t) \sim t^{1/4}$ \cite{MAZENKOCOP1} and included an
ad-hoc term to enforce $L(t) \sim t^{1/3}$ behavior.
The modified version still contains unphysical features
\cite{MAZENKOCOP2} such as the violation of the Tomita sum rule
\cite{TOMITA,SHINOZAKIOONO}.

In this paper we will concentrate on phase ordering dynamics without
conservation of order parameter but we will also discuss the case
in which the order parameter is conserved.
For the non-conserved order parameter
case, the closure schemes
reproduce many features observed in domain growth, such as dynamical
scaling and predictions for the growth of $L(t)$, the structure factor
\cite{OHTA82,OONO88a,MAZENKO} and the decay of two-time correlations
\cite{YEUNG90a,YEUNG90b,LIUAUTO}.  Variations of these models have been
used to study finite-size effects \cite{GUO}, finite temperature
effects \cite{OONO88a,GRANT}, effects of long-range initial conditions
\cite{HUMAYUN}, systems with long-range interactions \cite{HAYAKAWA},
and systems with a higher component order parameter
\cite{TOYOKI,BRAYANDPURI,LIUN2}.  However, the predictions of the
different closure schemes do not always coincide \cite{LIUDIM}.

In Section II, we derive a modified form of the OJK result
starting with the interface approach
(scheme I) correcting for inconsistencies in the original argument
\cite{OHTA82}.  In Section III, we discuss the bulk $u$-closure (scheme
B$u$) for the NCOP case and demonstrate that asymptotically it
provides the same result as the interface approach (scheme I). This
was originally suggested by Oono and Puri using a heuristic argument
\cite{OONO88a}.  We next discuss the bulk $\psi$-closure (scheme
B$\psi$).  As shown by Liu and Mazenko \cite{LIUDIM}, this closure
scheme leads to different predictions from the interface approach.
We will discuss each approach in a parallel manner to emphasize the
relation between the approaches.  We will show that the B$u$ and
B$\psi$ closures uses the same approximation so that their different
predictions signify the breakdown of that approximation.  In Section IV
we study the approximate Gaussian nature of $u( {\bf r}, t )$ through
numerical simulations.  We find that the single point probability
distribution function $P(u)$ decays as a Gaussian at the tails but
decays slower than Gaussian near $u = 0$.  In Section V we present
our findings for the conserved order parameter case.  We find that, in
this case, the Gaussian closure is fundamentally flawed.  In Section VI
we summarize our findings.

\setcounter{equation}{0}
\section{Interface Approach}

In this section, we derive a modified form of the Ohta-Jasnow-Kawasaki
and Oono-Puri results demonstrating the features of the
interface scheme.
We assume the dynamics
are described by the
time-dependent Ginzburg-Landau (TDGL) equation
\begin{equation}
	\frac{ \partial \psi }{ \partial t}
		=
	-\mu_{B}(\psi) + \frac{ \xi^{2}}{2} \nabla^{2} \psi,
\label{TDGLeq}
\end{equation}
where $\xi$ is the interfacial width, $\psi( {\bf r}, t )$ is the
scalar order parameter and $\mu_{B}(\psi)$ is the portion of the local
chemical potential which contains no gradient terms.
We assume
$\mu_{B}$ is an odd function of $\psi$, the equilibrium values of
$\psi$ are $\pm 1$; and there exists a
solution, $\psi = f(z)$, corresponding to a stationary planar interface
at $z = 0$, {\em  i.e.}, $f(z)$ obeys
$$
	0 = -\mu_{B}(f(z) ) + \frac{ \xi^{2}}{2} \frac{ d^{2} f }{d z^{2} }.
$$
For example, if $\mu_{B} = -\psi + \psi^{3}$ then $f(z) = \tanh(z/\xi)$.
However, other than the requirement that $f(z)$ increases monotonically
from $-1$ to $1$ over a length-scale $\xi$, the exact form $f(z)$ is
unimportant \cite{OONO88a,OONO88b}.
This universality is closely related to the universality in the dispersion
relation around the interface.\cite{chdisp}

At late times, the width $\xi$ is small relative to the characteristic
domain size $L(t)$, and the domain growth is determined by the motion
of the sharp interfaces.  The interface dynamics can be derived from
the TDGL \cite{ALLEN,KAWASAKI82,BRONSARD},
\begin{equation}
	v_{n}( {\bf r}, t ) = -\frac{ \xi^{2} }{ 2 } \; \kappa( {\bf r}, t ).
\label{INTEReq}
\end{equation}
Here the normal velocity of the interface, $v_{n}$, is defined as
positive when the `minus' phase moves into the `plus' phase, the normal
${\bf \hat{n}}$ points into the `plus' phase; and the local curvature,
$\kappa = \nabla \cdot {\bf \hat{n}}$, is positive for a bump of the
`minus' phase into the `plus' phase.

The essence of the interface method is to rewrite the interfacial
equation in terms of an indicator field $u( {\bf r},t)$.  The indicator
field is defined so that $u > 0$ ($u < 0$) in the plus (minus) phase,
$u = 0$ at the interface, and, near the interface, $|u|$ is the
distance to the interface, so that, $\nabla u({\bf r}, t) = {\bf
\hat{n}}({\bf r},t)$ near $u = 0$ \cite{OHTA82,OONO88a}. The main
motivation for introducing $u({\bf r},t)$ is that it remain continuous
on all length-scales, while $\psi({\bf r},t)$ is effectively
discontinuous upon rescaling distances by $L(t)$.  Therefore simple
decoupling approximations should be more trustworthy for the indicator
field $u$ than for order parameter field $\psi$.

In terms of $u$ the interfacial dynamics (Eq.\ (\ref{INTEReq}))
is
\begin{equation}
	\frac{ \partial u}{\partial t}
	=
	\frac{\xi^{2}}{2 } \nabla^{2} u.
\label{INTERUeq}
\end{equation}
However, Eq.\ (\ref{INTERUeq})
can only hold at $u = 0$, since, if it were true in
the bulk, the condition that $| \nabla u | =
1$ near the interface would be violated \cite{OONO88a}.  We
assume that the $| \nabla u |$ condition can be met by extending
Eq.\ (\ref{INTERUeq}) into the bulk by adding a  ``Lagrange
multiplier function''
$\tilde{P}(u,\nabla u) = P(u,\nabla u) u$ to the RHS of
Eq.\ (\ref{INTERUeq}),
\begin{equation}
	\frac{ \partial u}{\partial t}
	=
	\frac{\xi^{2}}{2}
	\left[
	\nabla^{2} u
	+
	 P( u, \nabla u ) u \right].
\label{BULKeq}
\end{equation}
The function $\tilde{P}$ has the following properties:
$\tilde{P}(0,\nabla u) = 0$, since Eq.\ (\ref{INTERUeq}) must be
recovered at $u = 0$, Due to the symmetry of the TDGL,
$\tilde{P}(u,\nabla u)$ is an odd function of $u$ and an isotropic
function of derivatives of $u$. Since the interface dynamics depends
only on local properties of the interface, $\tilde{P}(u,\nabla u)$ is
assumed to be local, {\em i.e.}, $\tilde{P}(u,\nabla u)$ depends only
on a finite number of derivatives of $u$.  Using heuristic arguments,
Oono and Puri replaced $P(u,\nabla u)$ by a function of time chosen to
maintain the equilibrium interface thickness \cite{OONO88a}. In the
following section, we demonstrate that $P(u,\nabla u)$ can be obtained
exactly from the TDGL using the bulk $u$-closure (B$u$ approach) and
therefore schemes I and B$u$ are asymptotically equivalent for the
non-conserved order parameter case.

 From Eq.\ (\ref{BULKeq}), the two-point correlation function, $\langle
u_{1} u_{2} \rangle$, obeys
\begin{equation}
	\frac{ \partial \langle u_{1} u_{2} \rangle }{ \partial t_{1} }
	=
	\frac{\xi^{2}}{2 }
	\left( \nabla_{1}^{2} \langle u_{1} u_{2} \rangle
	+
	\langle
	P( u_{1}, \nabla_{1} u_{1} ) u_{2} \rangle \right),
\label{BULKCORReq}
\end{equation}
where $u_{i} = u( {\bf r}_{i}, t_{i} )$ and $t_{1}
\neq t_{2}$.  Note that the local constraint $|\nabla u| = 1$ at
the interface forces $\langle \, u_{1}^{2} \, \rangle$ to grow as $L_{1}^{2}$,
where $L_{i} = L(t_{i})$.

We now need to make an assumption concerning the statistics of
$\{ u( {\bf r}, t) \}$.
The simplest assumption is that $u$ is a Gaussian stochastic
field.
As noted previously, the rationale is that $u$ is `equicontinuous' for
all length scales, so simple decoupling scheme may be more trustworthy
for it than for the original order parameter field.

For any  Euclidean symmetric Gaussian stochastic field, $\langle P(
u_{1},\nabla u_{1}) \, u_{1} u_{2} \rangle$ has a simple form
\begin{equation}
	\langle P( u_{1},\nabla u_{1}) \, u_{1} u_{2} \rangle
	=  p( \langle \;
u_{1}^{2} \; \rangle, \nabla^{2} \langle \, u_{1} u_{1} \, \rangle )
	\; \langle \, u_{1} u_{2} \, \rangle,
\end{equation}
where $\nabla^{2} \langle \, u_{1} u_{1} \, \rangle =
\lim_{{\bf r}' \rightarrow
{\bf r}_{1} } \nabla^{2} \langle u( {\bf r}_{1}, t_{1} ) u( {\bf r}',
t_{1} ) \rangle$.  Note that the function $p$ does not depend on
$t_{2}$ or ${\bf r}_{1} - {\bf r}_{2}$ so we can
replace $p$ as a function of $t_{1}$:
\begin{equation}
	\langle P( u_{1},\nabla u_{1}) \, u_{1} u_{2} \rangle
	=  p(t_{1}) \, \langle \, u_{1} u_{2} \, \rangle.
\end{equation}
For $t_{1} \neq t_{2}$ Eq.\ (\ref{BULKCORReq}) becomes
\begin{equation}
	\frac{ \partial \langle u_{1} u_{2} \rangle }{ \partial t_{1} }
	=
	\frac{\xi^{2} }{2}
	\left( \nabla_{1}^{2} \langle u_{1} u_{2} \rangle
	+
	p( t_{1} ) \,
	\langle u_{1} u_{2} \rangle \right).
\label{OJKOPeq}
\end{equation}
For $t_{1} = t_{2} = t$, we have
\begin{equation}
	\frac{ \partial g(r,t) }{ \partial t }
	=
	\xi^{2}
	\left( \nabla^{2} g(r,t)
	+
	p(t) \,
	g(r,t) \right),
\label{EQTIMEeq}
\end{equation}
where $g(r,t) \equiv \langle u(r,t) u(0,t) \rangle$.  The Gaussian
approximation means that the detailed local constraint $|\nabla u| = 1$
is no longer met but is replaced by a global constraint $\langle \, u^{2} \,
\rangle \sim L^{2}$.  This  requires that $ \xi^{2} p(t) = (d+2)/( 2 t)$,
where $d$ is the spatial dimension. This is exactly the form of $p(t)$
chosen in \cite{OONO88a} so that the interface thickness is time
independent.

The final step is to obtain the correlation function of $\psi({\bf r},
t)$ from $u({\bf r}, t)$.  Near the interface $\psi$ will be close to
the planar interfacial profile so that $\psi({\bf r}, t) = f(u({\bf r},
t))$.  At long times, the exact form of $f$ should be irrelevant, (as
long as $f$ approaches
$\mbox{sgn}(u)$ as $\xi \rightarrow 0$) \cite{SHINOZAKIOONO}.
Oono and Puri chose a form that simplifies the Gaussian integrals,
\begin{equation}
	f(u) = \mbox{sgn}(u) \ast \frac{1}{ (2 \pi \xi^{2})^{1/2} }
		\exp\left( - \frac{ u^{2} }{ 2 \xi^{2} } \right)
\label{SIMPLEFeq}
\end{equation}
where $\ast$ is the convolution.  They find
\begin{equation}
	C_{12}(r)
	=
	\frac{2 }{\pi}
	\arcsin\left( \frac{ \langle \, u_{1} u_{2} \, \rangle }{
	( \langle \, u_{1}^{2} \, \rangle + \xi^{2} )^{1/2}
	( \langle \, u_{2}^{2} \, \rangle + \xi^{2} )^{1/2} } \right),
\label{PSIPSIeq}
\end{equation}
where $C_{12}( r ) \equiv \langle \, \psi_{1} \psi_{2} \, \rangle$
and $r = | {\bf
r}_{1} - {\bf r}_{2} |$.  For any reasonable form of $f(u)$, a WKB-type
argument can be used to show that, $C_{12}(r)$ approaches
Eq.\ (\ref{PSIPSIeq}) when $\langle \,  u_{i}^{2} \, \rangle \gg \xi^{2}$.

In the original OJK analysis (as well as OP) there is a factor
$(d-1)/d$ in front of the diffusive term in the evolution equation for
$u$.
This difference is due to the lack of local constraint, $|\nabla u|
=1$, in OJK.   (For OP, this is due to both its eclectic and logically
opaque nature and the lack of this exact constraint.)
With this exception, Eq.\ (\ref{EQTIMEeq}) with $p(t)
= (d+2)/( 2 \xi^{2}  t)$ together with Eq.\ (\ref{PSIPSIeq}) is the
main result in Ref.\ \cite{OONO88a}.  The result of OJK can be obtained
from this approach asymptotically  without the difficulties in the
original argument.  In the original OJK derivation $u$ is not
self-consistently defined, since $f$ was chosen to be $\mbox{sgn}(u)$
and
$ |\nabla u| = 1$ was not even approximately enforced
\cite{OHTA82,OONO88a}.)

The interface approach predicts the following \cite{OHTA82,OONO88a}:
(1) dynamical scaling with $L \sim t^{1/2}$, (2) the equal-time
correlation function, $C(r,t)$ decays as a Gaussian at large $r$, (3)
the scattering function, $S_{k}$, displays Porod's law, $S_{k} \sim
k^{-(d+1)}$ \cite{POROD,TOMITA}, for the wavevectors $k$ in the
range $L^{-1} \ll k \ll \xi^{-1}$ and (4) Tomita's sum rule is obeyed
\cite{TOMITA}, {\em i.e.}, in the scaling limit, $\partial^{m} C/
\partial x^{m} |_{x = r/L \rightarrow 0^{+}} = 0$ for any even $m$.
For the two-time behavior, the interface approach predicts $C_{12}(0)
\sim (t_{1}/t_{2})^{-d/2}$ for $t_{1}/t_{2} \ll 1$ and the
autocorrelation function of the Fourier transforms of $\psi$ decays as
a stretched exponential, $\ln \langle \, \psi_{{\bf k}}(t_{1})
\psi_{-{\bf k}}(t_{2}) \, \rangle \sim -(t_{2}/t_{1})^{1/2}$, for $L_{2} \gg
L_{1}$ \cite{YEUNG90b}.

An alternative closure method is the one due to Kawasaki, Yalabik and
Gunton \cite{KAWASAKI78}.  This method tries to sum a diverging series
whose $n$-th term is of order $\exp( n\gamma t )$ with $\gamma t \gg
1$.  Although the result is finite and the scaling results are the same
as that of Eq.\ (\ref{EQTIMEeq}), this diverging behavior causes the
ratio $\xi/L(t)$ to vanish exponentially fast, an absurdity caused
perhaps by the strongly diverging nature of the series.

\setcounter{equation}{0}
\section{Bulk Approach}

\subsection{Bulk $u$-closure (Scheme B$u$)}

In this subsection, we discuss the bulk $u$-closure
(B$u$).  We demonstrate that the results of the
interface approach
can be obtained
without the intermediate step of the interface description.

We introduce an auxiliary field $u({\bf r}, t)$ by
\begin{equation}
	\psi( {\bf r}, t ) = f( u( {\bf r}, t ) ).
\label{MAPPINGeq}
\end{equation}
The planar interface solution, $f$, obeys
$$
	\mu_{B}(f) = \frac{ \xi^{2} }{ 2 }
	\frac{ d^{2} f}{ d u^{2} }.
$$
At late stages, the interfacial profiles will be very close to the
planar profile. Therefore this nonlinear mapping is exactly of the form
required to enforce $u = 0$ at the interface and $ \mbox{\boldmath $\nabla$} u
= {\bf
\hat{n}}$ near the interface.

In terms of $f(u)$, the TDGL is
\begin{equation}
	\frac{ \partial f }{ \partial t }
	=
	-\frac{  \xi^{2} }{2}
	\left[ \frac{ d^{2} f }{ d u^{2} }
	- \nabla^{2} f \right].
\label{TDGLF0eq}
\end{equation}
Applying the chain rule for $df/dt$ and $\nabla^{2} f$, we
obtain an expression for $\partial u/\partial t$,
\begin{equation}
	\frac{ \partial u}{ \partial t }
	=
	\frac{ \xi^{2}}{2}
	\left[
	\nabla^{2} u
	+
	\left( 1 - |\nabla u|^{2} \right) Q(u) \right],
\label{TDGLUeq}
\end{equation}
where $Q(u) \equiv -(df/du)^{-1} d^{2} f/du^{2}$.  For $\mu_{B} = -
\psi + \psi^{3}$ we find $f(u) = \tanh(u/\xi)$ and $Q(u) = 2 f(u)$.  In
general, $Q(0) = 0$, $Q(u)$ is an odd function of $u$ and
$Q(u)$ approaches $d \mu_{B} / d \psi |_{\psi_{eq}}$
(a finite constant) as
$u \rightarrow \infty$.  The result is that $Q(u)$ must be proportional
to $\mbox{sgn}(u)$ in the limit $\xi \rightarrow 0$.

Equation (\ref{TDGLUeq}) is exactly the form required to
extend the interface equation into the bulk  using
the interface approach (Eq.\ (\ref{BULKeq})) of the previous section.
We can now
identify the previously unknown `Lagrange multiplier'
function, $\tilde{P}(u,\nabla u)$ in
Eq.\ (\ref{BULKeq}), with  $( 1 -
|\nabla u|^{2} ) Q(u)$.  We repeat the steps of the interface
approach with the difference that we now have an explicit
expression for $\tilde{P}(u,\nabla u)$

 From Eq.\ (\ref{BULKeq}) we obtain the expression for the correlation
function
$\langle u_{1} u_{2} \rangle$,
\begin{equation}
	\frac{ \partial \langle u_{1} u_{2} \rangle }{ \partial t }
	=
	 \xi^{2}
	\left[
	\nabla^{2} \langle u_{1} u_{2} \rangle
	+ \langle
	( 1 - |\nabla_{1} u_{1}|^{2} ) Q(u_{1}) u_{2} \rangle
	\right],
	\label{qeq}
\end{equation}
where we restrict the discussion to $t_{1} = t_{2} = t$.  The
more general case is easily obtained.

To this point there are no approximations.  To proceed further we again
assume that $\{ u({\bf r}, t ) \}$ is a Gaussian stochastic field.
Just as in scheme I, this assumption is totally uncontrolled.  In the
same spirit as before, we assume that explicit forms of $f$ and $Q$ are
unimportant as long as they approach $\mbox{sgn}(u)$ and $2
\mbox{sgn}(u)$ in the limit of $\xi \rightarrow 0$.  Therefore we
choose a form of $Q$ which simplifies the Gaussian integrals,
\begin{equation}
Q(u) = \frac{i}{\pi} \int_{-\infty}^{+\infty} d \omega \frac{1}{\omega}
\exp \left( - \frac{ \xi^{2} \omega^{2} }{2}- i \omega u \right).
\end{equation}
That is, $Q(u)$ is the sign function mollified by a Gaussian function.
Equation (\ref{qeq})
can now be computed explicitly as
\begin{equation}
	\frac{ \partial g({\bf r},t)}{\partial t} =
	 \xi^{2}
	\left[ \nabla^{2} g({\bf r},t) + \frac{1 + \nabla^{2}
	g(0,t)}{\sqrt{ \pi ( \xi^{2} + g(0,t))/2} } g({\bf r},t) \right],
\label{BULKUeq}
\end{equation}
where, as in the previous section, $g( {\bf r}, t ) =
\langle \; u_{1} u_{2} \; \rangle$ for $t_{1} = t_{2} = t$.
That is, we have an equation exactly the same form as Eq.\ (\ref{EQTIMEeq})
\begin{equation}
\frac{ \partial g({\bf r},t)}{ \partial t} =
	 \xi^{2}
	\left[ \nabla^{2} g({\bf r},t) + p(t) g({\bf r},t) \right],  \label{gw}
\end{equation}
where
\begin{equation}
	p(t)
	=
	\frac{1 + \nabla^{2}
	g(0,t)}{\sqrt{ \pi ( \xi^{2} + g(0,t))/2} }
\label{PTequation}
\end{equation}
is now an explicitly known function of $t$.

We are left with a closed equation for $g({\bf r}, t)$ which can be
solved as an initial value problem.  However, it is more instructive to
consider the scaling limit.  We define the characteristic length-scale
by $L^{2} \equiv g(0,t)/(-\partial^{2} g( r, t)/ \partial r^{2} |_{r
\rightarrow 0} )$ so that $\nabla^{2} g(0,t) = -d g(0,t)/L^{2}$.  With
this definition we can show that $L^{2} \approx 2  \xi^{2} t$.
Substituting this in Eq.\ (\ref{gw}) we find
\begin{equation}
\frac{d g(0,t)}{d t} =
-\frac{d}{2 t} g(0,t) +  \xi^{2} p(t) g(0,t),
\end{equation}
with $p$ now given by
\begin{equation}
	\xi^{2} p(t) =
	\frac{  \xi^{2}  - d g(0,t)/(2t)}{\sqrt{\pi( \xi^{2} + g(0,t))/2}}.
\end{equation}
The only meaningful asymptotic solution of this equation is $g(0,t)
\simeq 2  \xi^{2} t/ d + {\cal O}(1)$.  This implies that
asymptotically $ \xi^{2} p(t) = (d+2)/(2t)$, {\em i.e.,} the same
$p(t)$ as that given by the interface approach.  We also find that this
is the asymptotic form if we solve the initial value problem using the
explicit expression (Eq.\ (\ref{PTequation})) for $p(t)$.

Therefore we have demonstrated that the results of the interface
approach can be obtained directly from the time-dependent
Ginzburg-Landau equation without explicitly using the interface
description.  In particular we have shown that the previously unknown
functions, whose properties were obtained by scaling arguments and
intuitive physical requirements, can be obtained more directly and,
furthermore, is exactly the form required by the scaling arguments.

\subsection{Bulk $\psi$-closure (Scheme B$\psi$)}

In the bulk $\psi$-closure the auxiliary field $u( {\bf r}, t)$ is
introduced in the same way as in the bulk $u$-closure, {\em i.e.},
$\psi({\bf r}, t) = f( u( {\bf r}, t ) )$, where $f$ is the planar
interfacial profile.  A dynamical equation is then obtained directly
for $\langle \; \psi( {\bf r}_{1}, t_{1} ) \; \psi( {\bf r}_{2}, t_{2} )
\; \rangle$
under the assumption that $\{ u( {\bf r}, t ) \}$ is a
Gaussian random field.  Mazenko applied this scheme to the non-conserved
order parameter case \cite{MAZENKO}.  We summarize the argument below.

The TDGL equation in terms of $\psi = f(u)$ is given by
Eq.\ (\ref{TDGLF0eq}),
\begin{equation}
	\frac{ \partial f }{ \partial t }
	=
	-\frac{  \xi^{2} }{2}
	\left( \frac{ d^{2} f }{ d u^{2} }
	- \nabla^{2} f \right).
\label{TDGLFeq}
\end{equation}
The expression for the correlation function  $\langle \, \psi_{1} \,
\psi_{2} \, \rangle = \langle \, f_{1} \, f_{2} \, \rangle$ becomes
\begin{equation}
	\frac{ \partial \langle \, f_{1} f_{2} \, \rangle }{ \partial t }
	=
	 \xi^{2} \left(
		\langle \, \frac{ d^{2} f_{1} }{ d u_{1}^{2} }
		f_{2} \, \rangle
		-
	\nabla^{2} \langle \, f_{1} f_{2} \, \rangle \right),
\label{PSICORReq}
\end{equation}
where $f_{i} = f( u( {\bf r}_{i}, t_{i} ) )$.  Here, we again let $t_{1} =
t_{2} = t$ for simplicity.  An analogous equation can be written for
$t_{1} \neq t_{2}$.

To this point there has been no approximations.  To proceed further,
$u$ is again assume to be a Gaussian stochastic field.  Using the
properties of Gaussian variables, one finds $\langle \, (d^{2} f_{1}/d
u_{1}^{2}) u_{2} \, \rangle$ = $\partial \langle \, f_{1} f_{2} \, \rangle
 / \partial
\langle u^{2} \rangle$.  This gives a closed equation for $C(r,t) = \langle
\psi(r,t) \psi(0,t) \rangle$.  The equation can then be solved numerically
as an initial value problem.  For later times, the exact form of $f$ is
irrelevant and we can use the $f$ given in Eq.\ ({\ref{SIMPLEFeq}).
Performing the Gaussian integrals, we find, $\partial C / \partial
\langle \, u^{2} \, \rangle =
- 2\tan( \pi C/2)/2/ (\pi (\langle \, u^{2} \, \rangle + \xi^{2}
) )$, and
\begin{equation}
	\frac{1}{  \xi^{2} }\frac{ \partial C(r,t) }{ \partial t }
	=
	\frac{2}{\pi (\langle \, u^{2}} + \xi^{2}) \, \rangle
	\tan\left( \frac{ \pi
	C(r,t) }{2} \right)
	+
	\nabla^{2} C(r,t). \label{NCOPCLeq}
\end{equation}
For $\langle u^{2} \rangle \gg \xi^{2}$ a scaling form can be assumed.
A nonlinear eigenvalue problem is then found which can be solved
numerically to obtain the correlation function $C(r,t)$
\cite{MAZENKO}.

This closure yields \cite{MAZENKO}: (1) dynamical scaling with $L \sim
t^{1/2}$, (2) that the quasi-elastic scattering intensity obeys Porod's
law at large $k$, (3) that Tomita's sum rule is obeyed \cite{TOMITA},
and (4) that the real space correlation function decays slightly slower
than a Gaussian with $C(x) \sim x^{a} \exp( -b x^{2} )$ where $a > 0$
depends on $\lambda$ and, hence, $d$.  For the two-time behavior, Liu
and Mazenko predicts that $C_{12}(0) \sim (L_{1}/L_{2})^{b}$ for $L_{1}
\ll L_{2}$ where $b \approx 1.28$ for $d=2$ and $b = 1.63$ for $d=3$
\cite{LIUAUTO}.  This closure also predicts that the auto-correlation
function for the Fourier transforms will decay exponentially in
$t_{2}/t_{1}$ for $t_{2} \gg t_{1}$ \cite{LIUAUTO}.

\subsection{The relation between the different approaches}

 From the above discussion the bulk $u$-closure and interface approaches
are equivalent.  We also see that the B$u$-closure and the
B$\psi$-closure uses exactly the same assumption: $\psi = f(u)$ with
$u$ assumed to be Gaussian.  However, we also see that there are
discrepancies in the predictions of the two approaches.  These
discrepancies can be regarded as an indicator of the limit of the
reliability of the Gaussian assumption.  Experimentally \cite{COLLINS}
there is general consensus with regards to the features in which both
approaches agree, i.e., $L \sim t^{1/2}$, Porod's law \cite{POROD}, and
Tomita's sum rule \cite{TOMITA},
although these are kinematic consequences.

To investigate the discrepancies between the different models, one must
rely on numerical studies, either kinetic Ising model simulations with
Glauber dynamics or cell dynamical schemes (CDS) corresponding to the
time-dependent Ginzburg-Landau equation \cite{OONO88b,review}.  Humayun and
Bray \cite{HUMAYUN},
and Shinozaki \cite{ASunpub} find that the equal time correlation
function $C(x)$ obtained by a CDS is better fitted by
the OJK result, although, $C(x)$ decays faster than both predictions.
Liu and Mazenko studied the autocorrelation function $C_{12}(0)$ and
found that the decay exponent $b$ is approximately $1.25$ for $d=2$ and
for $d = 3$, preliminary results indicate $b \approx 1.8$, {\em i,e.},
the decay is faster than the OJK prediction of $d/2$ and more in
agreement with the prediction of the B$\psi$ closure.  On the other
hand for smaller ratios of times $t_{2}/t_{1} \leq 10$, the OJK result
seems to give a very good zero parameter fit to $C_{12}(0)$
\cite{YEUNG90a}.  The numerical data for the autocorrelation of the
Fourier transform $\psi_{k}$ \cite{FURUKAWA89} has not yet been
sufficient to compare the results of the two closures.  In this case
the bulk $u$-closure predicts that $\langle \psi_{k}(t_{1})
\psi_{-k}(t_{2}) \rangle$ decays as a stretched exponential for $t_{2}
\gg t_{1}$ \cite{YEUNG90b} while the bulk $\psi$-closure predicts an
exponential decay \cite{LIUAUTO}.  Nevertheless for many features, the
Gaussian closures give reasonable agreement with experiment.  A broad
statement would be that, while B$\psi$ gives better fits to the
two-time behavior, schemes B$u$ and, hence, the scheme I, gives better
fits to the equal time behavior.

The relation between the closures is further clarified in the $d
\rightarrow \infty$ limit.  It was shown by Liu and Mazenko
\cite{LIUDIM} that the scaling functions obtained from OJK and the
$B\psi$ closure coincide in the limit of $d \rightarrow \infty$.  Here
we show more explicitly that the two closures are equivalent in this
limit.  In the limit of $t \rightarrow \infty$ $\langle \psi_{1}
\psi_{2} \rangle$ depends only on the ratio $\alpha(r,t) =
g(r,t)/g(0,t)$ for $t = t_{1} = t_{2}$.  The evolution equation for
$\alpha(r,t)$ in the B$u$ closure can be obtained from
Eq.\ (\ref{BULKUeq}).  In the limit of large $t$ this becomes
\begin{equation}
	\frac{ \partial \alpha(r,t) }{ \partial t }
		=
	\xi^{2} \left( \frac{\alpha(r,t) }{\langle u^{2} \rangle}
	+\nabla^{2} \alpha(r,t) \right)
\label{ALPHAUeq}
\end{equation}
For the B$\psi$ closure we can obtain the dynamical expression for
$\alpha(r,t)$  from
Eq.\ (\ref{NCOPCLeq}),
\begin{equation}
	\frac{ \partial \alpha(r,t) }{ \partial t }
		=
	\xi^{2} \left( \frac{ \alpha(r,t) }{ \langle u^{2} \rangle }
	+ \nabla^{2} \alpha(r,t)
	+ \alpha(r,t) \; \left| \nabla \alpha(r,t) \rangle \right|^{2} \right).
\label{ALPHAPSIeq}
\end{equation}
We can examine the $d \rightarrow \infty$ limit by rescaling lengths by
$x = r/\sqrt{d}$.
Eq.\ (\ref{ALPHAUeq}) becomes
\begin{equation}
	\frac{ \partial \alpha(x,t) }{ \partial t }
		=
	\xi^{2} \left( \frac{\alpha(x,t)}{\langle u^{2} \rangle}
	+ \frac{1}{ d }
	\frac{ \partial^{2} \alpha(x,t) }{ \partial x^{2} }
	+ \frac{ d - 1}{ d } \frac{ \partial \alpha(x,t) }{x \; \partial x}
 	\right).
\end{equation}
While Eq.\ (\ref{ALPHAPSIeq}) becomes
\begin{eqnarray}
	\frac{ \partial \alpha(x,t) }{ \partial t }
		& = &
	\xi^{2} \left( \frac{\alpha(x,t)}{\langle u^{2} \rangle}
	+ \frac{1}{ d }
	\frac{ \partial^{2} \alpha(x,t) }{ \partial x^{2} }
	+ \, \frac{ d - 1}{ d \;  x } \frac{ \partial \alpha(x,t) }{\partial x}
	+ \frac{ \alpha(x,t) }{ d }
	\left( \frac{ \partial_{x} \alpha }{ \partial t } \right)^{2}
	 \right).
\end{eqnarray}
Taking the limit of $d \rightarrow \infty$, we find that the equations
for $\alpha(r,t)$ become identical.  Although this is not a proof, it
does indicate that the Gaussian approximation may be valid in the limit
of $d \rightarrow \infty$.

\section{Numerical Results}

In the previous discussion we have shown that the $\{ u( {\bf r}, t) \}$
is not a Gaussian random field.  However, we have also argued that this
assumption may be a reasonable first approximation.  In this section
we study the statistics of $\{ u( {\bf r}, t ) \}$ directly through
a numerical updating of Eq.\ (\ref{BULKUeq}).  As noted previously, no
approximations are needed to proceed from the TDGL (Eq.\ (\ref{TDGLeq}))
to Eq. (\ref{TDGLUeq})
so this is equivalent to a simulation of the
TDGL equation.

For numerical efficiency we choose $Q(u) = 2$ if $u > 1$, $Q(u) = 2 \,
u$ if $1 \geq u \geq -1$ and $Q(u) = -2$ if $u < -1$. This effectively
approaches 2 sgn$(u)$ in the limit of $L(t) \gg 1$. We discretize the
system with mesh size $\delta x = 1.0$ and time steps $\delta t =
.05$.  To reduce lattice effects we used a sphericalized Laplacian as
described in Ref.\ \cite{OONO88b}.  For these large time and space
steps the update corresponds to a CDS
\cite{OONO88b}.  The simulation was performed on $800 \times 800$
lattices with periodic boundary conditions.  and repeated on $400
\times 400$ lattices to check for finite size effects.  The results for
$n=400$ and $n=800$ begins to deviate at about $t=400$ indicating that
the data for $n=800$ is not affected by size problems.  An average was
taken over 18 independent initial conditions for the larger lattice.

Figure 1 shows a plot of $L(t)^{2}$ vs. $\langle u(r,t)^{2} \rangle$
for $t=25,  50, 100, 200$ and $400$.  For $L(t)$ we use the inverse
interfacial density.  The line has a slope of unity.  We see that
$L(t)^{2} \sim \langle u(r,t)^{2} \rangle$ in agreement with the
arguments above.  We next calculate the single point probability
function $P(u,t)$ for each time.  Figure 2 shows $\ln P(u^{2}/\langle
u^{2} \rangle)$ vs. $u^{2}/\langle u^{2} \rangle$ for $t = 50, 100,
200$ and $400$.  It is clear that the probability distribution scales
during this time range.  We also observe that the tail of the
probability distribution decays in a Gaussian manner but the region
near $u = 0$ flatter than that for the Gaussian distribution.  Figure 3
shows the same data plotted with $-\ln\left( -\ln P(u^{2}/\langle u^{2}
\rangle ) \right)$ vs.\ $\ln u^{2}/\langle u^{2} \rangle$.  The
straight line has a slope of 2.  Figure 4 shows the flatness $\langle
u^{4} \rangle/( \langle u^{2} \rangle^{2} )$.  The times are the same
as that of Figure 1 plus the point at $t=0$.  The flatness is $3$ at
$t=0$ since the initial distribution is Gaussian.  For larger times we
find that flatness is somewhat less than that expected for a Gaussian
distribution.

 From our numerical result we find that
the single point probability distribution of $u$ is approximately
Gaussian at the tails.  However, the deviation from Gaussian behavior
near $u = 0$ is very important since the location of the interface is
at $u = 0$ which we have assumed controls the dynamics.
We also note that we have only looked at the single point distribution
function.  To test the full Gaussianness of $\{ u( {\bf r}, t) \}$ we
must look also at the two-point and two-time distributions.


\setcounter{equation}{0}
\section{Conserved order parameter case}

The bulk $\psi$ closure
(B$\psi$-scheme) approach was originally introduced to study
systems with conserved order parameter (spinodal decomposition).
Here we summarize our study of the conserved order parameter case emphasizing
the difficulties of this approach.

The first important
difficulty is that one cannot simply use the nonlinear mapping
$\psi({\bf r}, t ) = f( u( {\bf r}, t ) )$ to define the indicator
field. This is because the
condition that $| \psi( {\bf r}, t ) | <
\psi_{eq}$ is not met so that the mapping is not invertible.
This is easily seen since, for the conserved
case, local equilibrium near the interface requires that the deviation of
$\psi$ from its planar interface value is proportional
to the local curvature.   However
an indicator field can still be introduced using
$$
	\psi( {\bf r}, t ) =
	f( u( {\bf r}, t ) ) +  \phi ({\bf r}, t)
$$
where the $\phi$ field accounts for the deviation from the planar interfacial
profile
and the $u$ field has the same properties as in  the
nonconserved case.  One can then make the Gaussian assumption
for the $u$ field with the $\phi$ field coupled in such a way as to
enforce the $L(t) \sim t^{1/3}$ growth \cite{MAZENKOCOP2}.

However, even with this extension we find that the Gaussian assumption is
incompatible with conserved order parameter dynamics.  Since the extra field
$\phi$ is of order $1/L$ we can neglect its direct effects in the correlation
function in the scaling limit.
We note that Eq.\ (\ref{PSIPSIeq}), the relation between
$\langle \psi \psi \rangle$ and $\langle u u \rangle$,
$$
        C(r,t)
        =
        \frac{2 }{\pi}
        \arcsin\left( \frac{ \langle \, u_{1} u_{2} \, \rangle }{
         \langle \, u^{2} \, \rangle } \right),
$$
is independent of the dynamics and is true as long as $u({\bf r}, t)$ is
a Gaussian stochastic field.  Therefore, assuming $u$ is a Gaussian field,
we can invert this relation
to obtain $\langle \, u_{1} u_{2} \, \rangle/ \langle \, u^{2} \, \rangle$
from an
empirically obtained $C(r,t)$.  Figure 5 shows the spectral density
$\langle u_{k}(t) u_{-k}(t) \rangle$ obtained in this manner
using $C(r,t)$ from a very accurate
three dimensional
CDS simulation of spinodal decomposition
\cite{SHINOZAKIOONO93}.  We find that
the spectral density
$\langle u_{k}(t) u_{-k}(t) \rangle$
becomes significantly negative at small wavenumbers $q = k L(t) < 0.5$.
(The peak of $\langle \psi_{k}(t) \psi_{-k}(t) \rangle$ occurs
at approximately $q = 1$.)
Since
the spectral density must be positive definite, we can say that
the Gaussian assumption is not a reasonable description
of the large length-scale behavior.  In addition we find that
(inset of Figure 5) there is also a violation at positivity in the
very important range of
$q$ from approximately 1.5 to 4.  This corresponds to the structure at
wavenumbers just above that of the peak of the scattering intensity.
We can conclude that the bulk closure approaches are inherently flawed for
the conserved order parameter case.

Since interface approaches also use the Gaussian assumption and
the same relation
Eq.(\ref{PSIPSIeq}), they are also flawed.  However, Ohta and Nozaki's modest
success justifies a more systematic study of the interface approach.
This is especially the case since
the Ohta-Nozaki approach fixes the growth exponent
to be 1/3 by an intuitive but rather ad hoc manner
\cite{OHTAandNOZAKI}.
Here we briefly summarize our interface attempt and its limitation.

Our starting point is the interface dynamic equation in terms of the
$u$ field
\begin{equation}
	G * \delta(u) \partial_{t} u = \nabla^{2} u,  \label{copinter}
\end{equation}
where $G$ is the Laplacian Green's function, i.e., $G = - \nabla^{-2}$
\cite{OHTAandKAWASAKI}.
As in the
non-conserved case, this
equation is correct only at the interface.  We assume the following
bulk extension of the interface equation:
\begin{equation}
 	G * \delta(u) \partial_{t} u = J \nabla^{2} u + Q, \label{copinterex}
\end{equation}
where $J$ is a function of $u$ such that $J=1$ at the interface, and
$Q$ is a functional of $u$ and its derivatives up to the second order
with $Q = 0$ at the interface.  The extension is motivated by an analogous
idea behind Eq.\ (\ref{BULKeq}).  Applying the Laplacian to
Eq.\ (\ref{copinterex}), and assuming that $u$ is a Gaussian stochastic
field, we find (after some algebra)
\begin{eqnarray}
	\frac{1}{2} \frac{1}{\sqrt{2 \pi g(0,t)}} \partial_{t} g(r,t)
	& = & -R(t)
	\nabla^{4} g(r,t)  - \nabla^{2} Q(t) g(r,t)
	+ P(t), \label{clounk}
\end{eqnarray}
where $g(r,t) = \langle u(r,t) u(0,t) \rangle$ and $P,Q$ and $R$ are yet
unspecified functions of time only.  We now assume that there is a scaling
regime and determine the forms of $R(t)$, $Q(t)$ and $P(t)$ necessary for
a scaling solution to exist.  From the definition of $u$ we require
$\langle u^{2} \rangle \sim L^{2}$ for large times, so that, in the
scaling limit, $g(r,t)$ must be of
the form
\begin{equation}
	g(r,t) = L(t)^{2} g(r/L(t)), \label{SCALeq}
\end{equation}
where  $g(0)=1$.
Rescaling $x = r/L(t)$, we get Eq.\ (\ref{clounk}) with
Eq.\ (\ref{SCALeq}) as
\begin{eqnarray}
	\lambda \left( 2 g(x) - x \frac{ d g}{d x} \right)
	& = & - R(t) \nabla^{4} g(x) -
	L(t)^{2} \nabla^{2} Q(t) g(x)
	+  L(t)^{4} P(t) g(x),
\label{SCALINTeq}
\end{eqnarray}
where $\lambda = L'(t) L(t)^{2}/2 \sqrt{2 \pi}$ which must be
time-independent asymptotically.  The coefficients $R$,$L^{2}Q$ and
$L^{4}P$ must converge to a non-zero constant in the $t \rightarrow
\infty$ limit; if they diverge, we get physically absurd results, while
the same is true if these coefficients vanish. Hence, asymptotically
Eq.\ (\ref{SCALINTeq}) becomes
\begin{equation}
	\lambda( 2g(x) - x \frac{d g(x)}{d x}) + \nabla^{4}
	g(x) + A \nabla^{2} g(x) - B g(x) =0,
\label{copclas}
\end{equation}
where we have rescaled $\lambda$ to get rid of the numerical
coefficient in front of the double Laplacian, and $A$ and $B$ are
constant.  For $g(0)$ to be finite, $B = (2 +d) \lambda$ is required.
Hence, in the $k$-space, we arrive at
\begin{equation}
\lambda \frac{dg}{dk} = (- k^{3} + k) g, \label{clcl}
\end{equation}
where $A$ is scaled out, which must be positive and $\lambda$ remains
an unspecified constant.  The resultant
equation is very similar to the Ohta-Nozaki equation, and has the same
defect, although in our case
$L \sim t^{1/3}$ follows from our starting point of
the interface equation.

Given these caveats we fit the empirically obtained $C(r,t)$, using
the $g(r,t)$ obeying Eq.\ (\ref{clcl}).  We find that
$\lambda = 0.013$ gives the best fit to $C(r)$.   Figure 6 shows that
the fit is very good up to approximately the second zero of
$C(r)$.  This is further shown in the inset in which $(r/L)^{2} C(r)$
is plotted to show that goodness of the fit is not simply because
$C(r)$ is small.  As noted previously, the Gaussian assumption is
invalid for long length scales and the fit becomes increasingly
worse for larger $x$.  The same holds for the scattering intensity.
We find a very good fit for $q > 0.5$ but the conservation law is violated
due to the invalidity of the Gaussian field assumption at small $k$.

To conclude, we find that the Gaussian assumption can give
a nontrivial fit
to a limited range of length-scales for the correlation function and
wavenumbers for the scattering intensity.  However, given the
fundamental flaw in the Gaussian assumption at larger distances, we
feel that there is no point in making a more concerted effort \cite{Tomita}.

\section{Discussion and Summary}

We have discussed the relation between different closure approximations
for phase ordering without conservation of order parameter.  These
closure approximations are based on the assumption of an underlying
Gaussian stochastic field $u( {\bf r}, t) $.  We discuss two general
methods, the interface approach in which a dynamical equation for $u(
{\bf r}, t)$ is obtained from the interfacial dynamics and the bulk
approaches in which the $u( {\bf r}, t)$ is defined by the nonlinear
relation $\psi( {\bf r}, t ) = f( u( {\bf r}, t) )$, where $f$ is the
planar interfacial profile.  The bulk approach is further subdivided
into the bulk $u$-closure and bulk $\psi$-closure in which dynamical
equations are obtained for the correlation function of $u$ and $\psi$,
respectively.\cite{BRAYandHUMAYUN93}

We have derived a modified form of the original interface approach
\cite{OHTA82} and shown that it is completely equivalent to the
bulk $u$-closure.  We then have demonstrated that the only assumptions
of the bulk $\psi$-closure and bulk $u$-closure is that
$\{ u( {\bf r}, t ) \}$ is a Gaussian stochastic field.  Our conclusion is
that the discrepancies in the predictions of the
bulk $u$-closure and bulk $\psi$-closure is due to a breakdown of that
approximation.

We have also discussed the Gaussian closure for the conserved case.  We
have shown that the Gaussian approximation is more fundamentally
flawed in this case.  However, the interface closure approach still
leads to a nontrivial fit of the correlation function $C(r,t)$ and
scattering intensities $S_{k}(t)$ except at large scaled distances
$r/L(t)$ or at small scaled wavenumbers $k L(t)$.

\acknowledgements

We would like to thank Prof.\ David Jasnow and Dr.\
Timothy Rogers for helpful
discussions.  C.Y.\ is grateful to the National Science Foundation for
support under grant DMR 89-14621 and Natural Sciences and Engineering
Research Council of Canada.  Y.O.\ and A.S.\ gratefully acknowledge the
support of the National Science Foundation under grant DMR 90-15791.


\newpage

\begin{figure}
\caption{A log-log plot of $\langle u^{2} \rangle$ vs.\ $L(t)^{2}$
from the $800 \times 800$ simulations.
The solid line is a slope of 1.  The statistical uncertainties
are smaller than the symbol sizes.
$L(t)$ is the inverse interfacial
density.  We find that $\langle u^{2} \rangle$
grows as $L^{2}$ as predicted by the interface approach.
A fit to the form $\langle u^{2} \rangle = a L^{2 b} + c$ gives
$b = 1.03$.}
\end{figure}

\begin{figure}
\caption{A plot of the single point probability distribution
$\ln P( u^{2}/\langle u^{2} \rangle, t)$ versus
$u^{2}/\langle u^{2} \rangle$ for $t=50, 100, 200$ and $400$.
Representative error bars are shown.
This plot indicates that the tail of the distribution function
decays as a Gaussian but there is a regime for
$u^{2}/\langle u^{2} \rangle < .5$ which
decays slower than predicted by the tails.}
\end{figure}

\begin{figure}
\caption{The same data as in Fig.\ 2 plotted in the form
$-\ln( -\ln P( u^{2}/\langle u^{2} \rangle, t) )$ vs.
$\ln (u^{2}/\langle u^{2}\rangle)$.  The line has a slope of
$-1$ indicating the Gaussian nature of the tail.
}
\end{figure}

\begin{figure}
\caption{The flatness $\langle u^{4} \rangle/ \langle u^{2} \rangle^{2}$
versus time.  The distribution for the initial condition is Gaussian
so that at $t = 0$ the flatness is 3.  For larger times the flatness
is somewhat smaller than that of a Gaussian distribution.
}
\end{figure}

\begin{figure}
\caption{Plot of the spectral density $\langle u_{k}(t) u_{-k}(t) \rangle$ from
the $\langle \psi_{k}(t) \psi_{-k}(t) \rangle$ obtained under the Gaussian
assumption from the 3-d spinodal decomposition simulation (dashed line).  The
inset is a blow up of the spectral density for values of $q$ just above the
peak.  We find that there is a violation of positivity both $q < .5$ and $1.5 <
q < 4$, indicating that the Gaussian assumption is clearly invalid for the
conserved order parameter case.}
\end{figure}

\begin{figure}
\caption{Plot of the real space correlation function $C(r,t)$ from the 3-d CDS
simulations (dash line) versus the result using the Gaussian assumption
with the B$\psi$ scheme (solid line).  The inset shows $(r/L)^{2} C(r,t)$
for the same range of $r/L$.  We find very good fits up to the second zero in
the correlation function.}
\end{figure}


\begin{thebibliography}{99}
\bibitem{BRAYandHUMAYUN93}
We have received a preprint by
A.J.\ Bray and K.\ Humayun which discusses the B$u$ closure
and its equivalence with the OJK interface approach using
the extra assumption that $Q(u) \sim u$.
\end{thebibliography}
\end{document}